\documentclass[10pt,letterpaper]{article}
\usepackage{color}
\usepackage{graphicx}
\usepackage{amsmath}
\usepackage{esint}
\usepackage{subfig}
\usepackage{textcomp}
\usepackage{braket}
\usepackage{mathrsfs}
\usepackage{placeins}
\usepackage{authblk}
\usepackage{cite}
\usepackage{relsize}
\usepackage{amssymb}
\begin{document}
\title{Detecting the spatial quantum uncertainty\\ of bosonic systems}
  
\author[1,2,3]{Vanessa Chille}
\author[3]{Nicolas Treps}
\author[3]{Claude Fabre}
\author[1,2]{Gerd Leuchs}
\author[1,2]{Christoph Marquardt}
\author[1,2]{Andrea Aiello}

\affil[1]{Max Planck Institute for the Science of Light, G\"unther-Scharowsky-Str. 1/Bldg. 24, D-91058 Erlangen, Germany}
\affil[2]{Institute of Optics, Information and Photonics, University of Erlangen-Nuremberg, Staudtstr. 7/B2, D-91058 Erlangen, Germany}
\affil[3]{Laboratoire Kastler Brossel, Sorbonne Universit\'e - UPMC, ENS, Coll\`ege de France, CNRS; 4 place Jussieu, 75252 Paris, France}

\maketitle
\begin{abstract}
We present the quantum theory of the measurement of bosonic particles by multipixel detectors. For the sake of clarity, we specialize on beams of photons.  
We study the measurement of different spatial beam characteristics, as position and width. The limits of these measurements are set by the quantum nature of the light field. We investigate how both, detector imperfections and finite pixel size affect the photon counting distribution. An analytic theory for the discretized measurement scheme is derived. We discuss the results and compare them to the theory presented by Chille \textit{et al.} in ``Quantum uncertainty in the beam width of spatial optical modes,'' Opt. Express {\bf 23}, 32777 (2015), which investigates the beam width noise independently of the measurement system. Finally, we present numerical simulations which furnish realistic and promising predictions for possible experimental studies.
\end{abstract}


\section{Noise in spatial beam parameters}
Consider raindrops falling on a terrace covered by tiles. The probability that, in a given time interval, exactly $N$ drops hit a specific tile, is known to follow a Poisson distribution. An analogue situation occurs in our eyes when, under the condition of very low light intensity, only a few photons are detected by the arrangement of high-sensitive rods \cite{Teich1982}. In this case, the probability that $N$ photons excite a given rod will not be generically Poissonian, but will depend on the quantum state of light entering the eyes. A similar circumstance takes place in many quantum optics experiments when photons are detected by a multipixel CCD camera. In all these three cases, one considers the spatial statistical distribution of either raindrops or photons detected by a pixelated surface. We want to note that, of course, there is a fundamental difference in the description of classical raindrops and quantum particles \cite{Smith2007}.\\
In this work, we develop a quantum theory of bosonic particle measurements by multipixel detectors. Specifically, we determine how some spatial characteristics of a beam of bosonic particles (as, e.g., its central position and width) are affected by the quantum state of the beam. As an exemplary case, we treat in great detail multipixel detection of electromagnetic particles, namely photons, a topic that has received increasing attention in recent years  \cite{ReviewKolobov, QuantumImaging, Displacement, Resolution, Gabriel}. Within this context, it is well-known that quantum noise limits the precision of optical measurements \cite{Bachor, Zoller}. Thus, the reduction of the quantum noise is a natural goal to aim for. In order to do so, one has to develop a deeper understanding of its origin. 

Prior investigations have been carried out for many different spatial beam parameters and in many different manners. In the late nineties, the transverse distribution of intensity noise in the far field of semiconductor lasers has been studied in \cite{SpatialNoiseLasers1998}. The authors have shown that a large amount of the noise is actually present in the transverse modes that are nonlasing. The spatial distribution of squeezing generated by parametric down conversion, either in near field or far field, has also been investigated\cite{Lugiato}. 

Later, there have been a number of papers about the measurement of small transverse displacements of light beams\cite{Treps2003}. They investigated optimal measurements, as well as the quantum noise, setting a limit to the precision of these measurements. Moreover, they showed that it is possible to reduce the quantum noise limit by introducing squeezed light in appropriate spatial modes \cite{Displacement, Hsu2004, Barnett}. In particular, measurement of a TEM$_{00}$ beam displacement by means of a homodyne detection utilizing a TEM$_{10}$ mode as a local oscillator has been compared to the measurement with a split detector \cite{Delaubert2006}.  

Another example for an application for which the quantum noise in the spatial domain is important is microscopy. In particular super-resolution microscopy requires the ability of determining the location of optical point sources accurately. In \cite{Resolution, Tsang2015}, the associated quantum limits are derived and an enhancement of the accuracy by the use of squeezed light in appropriate modes is discussed.

The usual way to monitor the transverse distribution of light in an image is by using a CCD camera, i.e. a multipixel detector. The topic of detecting spatial parameters by a multipixel detector is approached in a fundamental way in \cite{QuNoiseMultipixel2005}. The origin of quantum noise for different measurements, i.e. for different linear combinations of the pixel's outputs, is investigated. Moreover, it is shown that it is possible to reduce the noise below the standard quantum limit by utilizing squeezed light in a particular spatial mode for preparing the laser beam.

In the present work, we approach the problem from a more practical point of view and present a theory that allows us to take experimental imperfections and technical conditions into account. We provide a complete theoretical description of spatial measurements by means of a pixelized detector. Our aim is to investigate the experimental feasibility of the measurement of beam parameters and their noise by state-of-the-art detectors. In particular, we investigate the noise in the measurement of the beam width and the beam position.

The standard theory of quantum photodetection gives the probability distribution for counting $N$ photons in a given time interval, say $t$ to $t+T$ \cite{Loudon, MandelWolf}. Conversely, the main goal of this work is determining the probability distribution for counting $N$ photons in a given limited area, say a specific pixel, of the detector surface.\\
According to elementary statistical considerations, such a probability distribution for a uniformly illuminated multipixel detector is expected to be a multinomial distribution, basically arising from distributing $N$ photons amongst $M$ pixels. We found that this educated guess holds true for single mode states of the electromagnetic field. Conversely, for multimode states, in particular entangled ones, the situation is expected to be more complicated \cite{InPreparation}. In addition, we find that the spatial mode profile of the light beam affects the distribution of quantum noise as well.

We start the presentation of our work by giving an introduction to the formalism that we use to describe the light field and the multipixel detector in Sec.\,\ref{SecLightField}. We then determine the probability distributions for the photon counts of the set of pixels for both an ideal and  an imperfect photo detector. We develop an analytic theory for the noise in the beam width and position based on the derived probability function and compare it to the continuous theory presented in \cite{BWU}, which is independent from the measurement system. Finally, the results of our numerical simulations are presented and the associated experimental measurement procedure is discussed. We take technical difficulties and imperfections into account and thus give the results for realistic measurement conditions. While \cite{BWU} provided a fundamental investigation of the quantum uncertainty present in the beam width, this work shows what is experimentally measurable, and thus how the noise affects realistic measurements.


\section{Description of the light field} \label{SecLightField}
A light field can be expanded in whatever mode basis one prefers. Usually, one chooses the basis that fits best to the light field one seeks to describe. Common bases are tilted plane waves, and, for the subset of paraxial beams propagating in a given mean direction, the Hermite-Gauss or the Laguerre-Gauss modes.   

More generally, we may choose to decompose the complex field of interest $E(\textbf{x},z)$ on any orthonormal set of normalisedsolutions of Maxwell equations $\{\Phi_K(\textbf{x},z)\}$ \cite{Deutsch1991, Blow1990}:
\begin{equation}
E(\textbf{x},z)=\sum_K E_K \Phi_K(\textbf{x},z),
\end{equation}
where $\textbf{x} = (x,y)$ are the transverse coordinates.  According to the chosen set of modes, $E(\textbf{x},z)$ can be any field solution of Maxwell's equations, or a subset of these. In the following, we assume that the light beam propagates along the $z$-axis, and we consider the transverse variation of fields only in the $(x,y)$ plane where the CCD camera is placed. Therefore we can omit the $z$ variation of the various quantities that we will introduce. The single index $K$ may actually denote a pair of independent indices. For example, for Hermite-Gauss modes, $K$ stands for $(n,m)$, with $n,m \in \{0,1,2,...\}$.

An annihilation operator $\hat{a}_K$ is associated with each mode. It is determined by integrating the local annihilation operator $\hat{a}(\textbf{x})$ over the mode $\Phi_K(\textbf{x})$ in the CCD camera plane:
\begin{align}
\hat{a}_K &= \int\text{d}^2\textbf{x} \Phi^*_K(\textbf{x})\hat{a}(\textbf{x}),\label{Eq2}
\end{align}
where, by definition, $\left[\hat{a}(\textbf{x}),\hat{a}^\dagger(\textbf{x}^\prime)\right] = \delta(\mathbf{x}-\mathbf{x}^\prime)$. 

If the set of modes $ \Phi_K$ is a basis of the whole space of Maxwell equations (for example the plane wave and the Hermite-Gauss basis), one has in addition the completeness relation
\begin{align}
\sum_K  \Phi_K^*(\textbf{x},z)\Phi_K(\textbf{x}^\prime,z)=\delta(\mathbf{x}-\mathbf{x}^\prime)
\end{align}
which enables us to invert relation (\ref{Eq2})
\begin{align}\label{ak}
\hat{a}(\textbf{x})= \sum_K \hat{a}_K \Phi_K(\textbf{x},z)
\end{align}

In other cases, it is beneficial to adapt the mode basis to the system the light is interacting with. As we are investigating a CCD camera with discrete pixels, we define operators $\hat{a}_{ij}$ associated with the pixels resembling the annihilation operators $\hat{a}_K$ for the modes $\Phi_K$ in Eq.\,\ref{Eq2}: the operator $\hat{a}(\mathbf{x},z)$ is integrated over the surface of the pixel. For this purpose, the two-dimensional step function $\Theta_{ij}(\mathbf{x}) = \Theta(d-2|x-x_i|)\Theta(d-2|y-y_i|)$ is utilized, where $\{i,j\}$ is the pair of coordinates characterizing the position of the individual pixel centered at $\mathbf{x}_{ij} = \{x_i,y_j\}$ and d is the pixel width. We thus write
\begin{align}
\hat{a}_{ij}(z) = \frac{1}{d} \int\text{d}^2\mathbf{x} \Theta_{ij}(\mathbf{x})\hat{a}(\mathbf{x},z), \label{aij}
\end{align}
where the factor $1/d$ ensures the normalization. Please see the end of this section for further comments on this approach. The resolution of the measurement is limited due to the integration over a finite region of transverse space. According to the Shannon-Nyquist sampling theorem \cite{Shannon1949}, the upper limit of the measurable spatial frequency components equals $1/2d$. This limitation is inherent in the measurement scheme, i.e. unavoidable when a multipixel detector is utilized as it is necessarily discretized.

\begin{figure}[h]
\centering
    \includegraphics[width=0.5\textwidth]{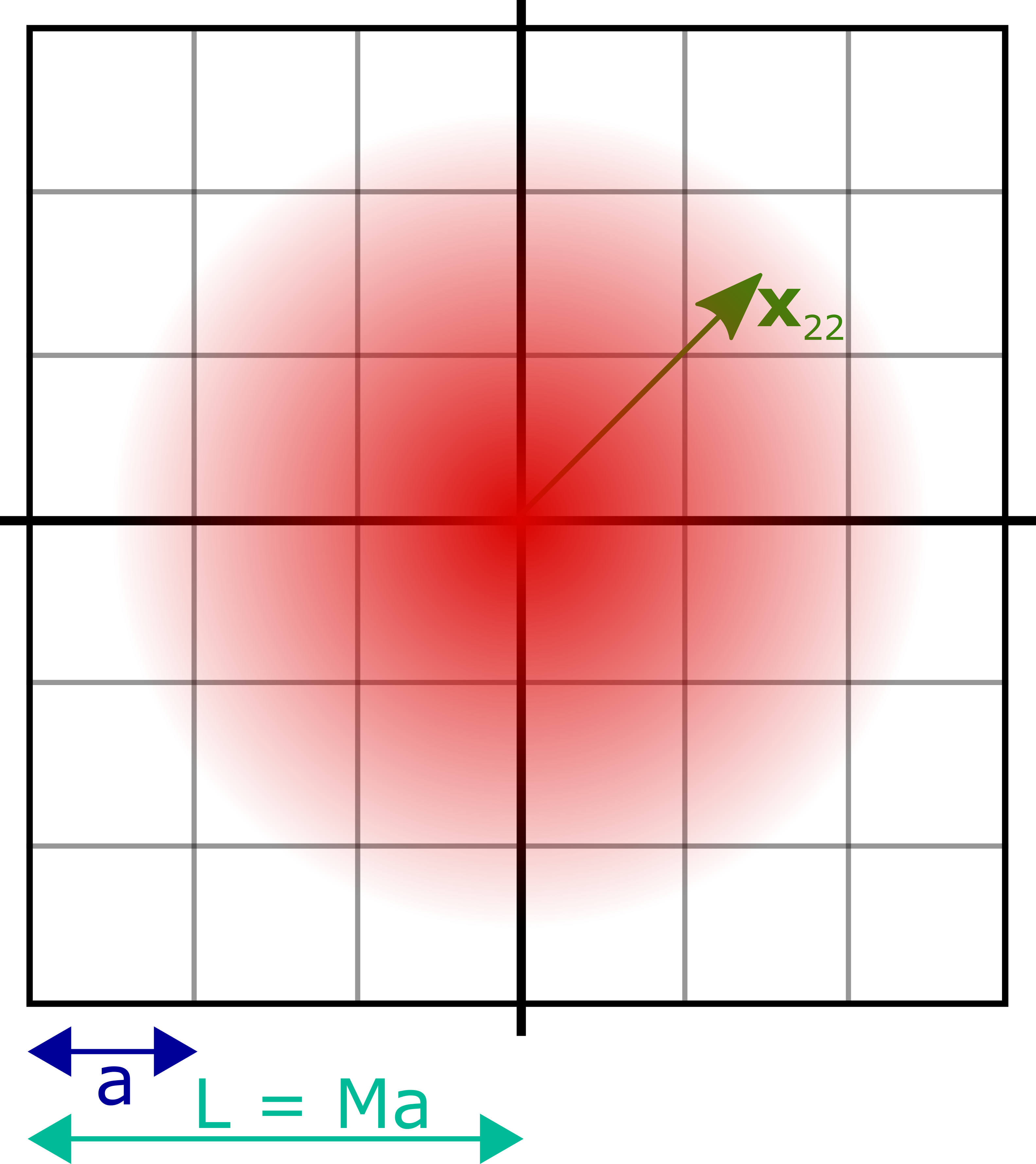}
		\caption{\textit{CCD camera.} The pixel array of the detector is illuminated by the impinging light beam, for example with the spatial profile of a fundamental Gaussian mode.}
    \label{fig:CCD}
\end{figure}

To clarify the nomenclature for the description of the CCD camera, the sketch in Fig.\,\ref{fig:CCD} gives a visualization. A fundamental Gaussian beam impinging on the set of pixels is depicted. We assume the pixels are squares of width $d$. In total, the camera possesses $N = 4M^2$ pixels, where $2L$ is the width of the square CCD camera, with $L = Md$. The indices $i$ and $j$ defining each pixel are chosen such that the $x$ and $y$ coordinates of the pixels' centers are given by

\begin{subequations}
\begin{align}
x_i = -L+\left(i-\frac{1}{2}\right)d,\\
y_j = -L-\left(j-\frac{1}{2}\right)d.
\end{align}
\end{subequations}
The top-left pixel is thus centered at $\textbf{x}_{11} = \left( -L+d/2, L-d/2 \right)$ and the bottom-right pixel is centered at $\textbf{x}_{2M,2M} = \left( L-d/2,-L+d/2 \right)$.\newline

To investigate the effect of the impinging light beam on the CCD camera, we perform the transition from the mode basis $\{\phi_K(\mathbf{x},z)\}$ to the pixel basis. We thus link the description that suits the light field to the description that matches the measurement system. We express $\hat{a}_{ij}$, in terms of the annihilation operators of the complete mode basis $\hat{a}_K$. Substituting Eq.\,(\ref{ak}) in Eq.\,(\ref{aij}), we obtain
\begin{align}
\hat{a}_{ij}(z) = \frac{1}{d} \sum_K \hat{a}_K \mathscr{U}_{Kij}(z),
\end{align}
where we have defined the mode-pixel superposition coefficient as
\begin{align}
\mathscr{U}_{Kij}(z) = \int\text{d}^2\mathbf{x} \Phi_K(\mathbf{x},z)\Theta_{ij}(\mathbf{x}).
\end{align}
We have thus obtained a very useful expression that allows us to write down the effect of the light modes on the pixels. For the moment, we assume that the multipixel detector counts all of the photons impinging on each pixel. Imperfect detection will be considered later. We are thus searching for the expression for a quantum state with $n$ photons localized within the pixel centered at $\textbf{x}_{ij}$. We can write this state by means of $\hat{a}_{ij}^\dagger$ for the pixel $(i,j)$ that we have just determined as
\begin{align}\label{Eq9}
\ket{n;ij} = \frac{1}{\sqrt{n!}}\left( \hat{a}_{ij}^\dagger \right)^n \ket{0}.
\end{align}
Following the same approach as in Eq.\,(\ref{Eq2}), we define a pixel photon number operator $\hat{n}_{ij}$ as
\begin{align}\label{Eq10}
\hat{n}_{ij}(z) = \int\text{d}^2\mathbf{x}\hat{a}^\dagger(\mathbf{x},z)\hat{a}(\mathbf{x},z)\Theta_{ij}(\mathbf{x})
\end{align}
such that
\begin{align}\label{eigenstaten}
\hat{n}_{ij} \ket{n^\prime;kl} = n^\prime \delta_{ik}\delta_{jl}\ket{n^\prime;kl}.
\end{align}
The following equal-$z$ canonical commutation relations hold:
\begin{align}
[\hat{a}(\mathbf{x},z),\hat{n}_{ij}(z)] &= \Theta_{ij}(\mathbf{x})\hat{a}(\mathbf{x},z),\\
[\hat{n}_{ij}(z),\hat{n}_{kl}(z)] &= 0,\\
[\hat{a}_{ij}(z),\hat{n}_{kl}(z)] &= \delta_{ik}\delta_{jl}\hat{a}_{ij}(z).
\end{align}
By means of the operator $\hat{n}_{ij}$, the photons measured by the pixel $(i,j)$ can be counted. The pixel annihilation operators $\hat{a}_{ij}(z)$ defined via Eq.\,(\ref{aij}), have been introduced with the aim of  building the eigenstates $\ket{n; ij}$ of the pixel photon number operator,
as shown in Eq.\,(\ref{Eq9}). The definition of $\hat{n}_{ij}(z)$ given in Eq.\,(\ref{Eq10}) is the most natural one: this operator simply counts the total number of photons recorded by the pixel $(i,j)$. However, the pixel annihilation and creation operators $\hat{a}_{ij}(z)$ and $\hat{a}_{ij}^\dagger(z)$, respectively, are not related to the pixel photon number operator $\hat{n}_{ij}(z)$ by the common relation $\hat{n}_{ij}= \hat{a}_{ij}^\dagger \hat{a}_{ij} $. This is due to the fact that although Eq.\,(\ref{Eq2}) and Eq.\,(\ref{aij}) are formally similar, their physical content is quite different. In the first case, the mode functions $\Phi_K(\mathbf{x})$ form a complete basis of $\mathbb{R}^2$; in the second case the ``pixel mode functions'' $(1/d) \Theta_{ij}(\mathbf{x})$, do not (at least, not in the conventional sense). In fact, we built the pixel mode functions exploiting the concept of Weyl's eigendifferentials, which are special vector states that have only a finite extension in space (in the  $\mathbb{R}^2$ domain), as illustrated in detail in \cite{Greiner1993, Halcomb1984, Messiah1961}.
\section{Probability distribution for the photons measured by an ideal photodetector}
To make a statement about spatial measurements and their noise, we need to know how the impinging photons are distributed over the surface of the detector. We thus determine the photon probability distribution in the following. This means that we have to find an answer to the question how likely it is to find a data set with $n_i$ photons in the $i$-th pixel. We start by investigating an ideal photodetector where the number of photons is equivalent to the number of counts and extend our theory to realistic, i.e. imperfect detectors, afterwards.\newline
We assume that the ideal multipixel detector is placed in a given plane perpendicular to the propagation axis $z$ and that the beam is prepared in the quantum state $\ket{\psi}$. The probability that the pixel centered at $\textbf{x}_{i_1 j_1}$ will count $n_1$ photons, the pixel centered at $\textbf{x}_{i_2 j_2}$ will count $n_2$ photons, ..., and the pixel centered at $\textbf{x}_{i_N j_N}$ will count $n_N$ photons can be expressed by the overlap of the impinging state $\ket{\psi}$ and the state with $n_i$ photons in the $i$-th pixel $\ket{n_1,n_2,...,n_N}$:
\begin{align}\label{EqP}
P(n_1,n_2,...,n_N|\psi) = |\braket{n_1,n_2,...,n_N|\psi}|^2.
\end{align}
In order to simplify the notation, we use $\ket{n_\nu}$ as shorthand for $\ket{n_\nu;i_\nu j_\nu}$ and $\hat{a}_\nu = \hat{a}_{i_\nu j_\nu}$ in the following. We define the cumulative index $\nu$ as $\nu = \nu(i,j) \equiv 2M(i-1)+j$. 

For the sake of simplicity, we will restrict our studies in the present paper to single mode states in a mode $\Phi_K(\textbf{x},z)$ of given $K$-value. $\ket{\psi}$ can then be written as
\begin{align}
\ket{\psi} = \sum_n \psi_n \ket{n} = \sum_n \frac{\psi_n}{\sqrt{n!}} \left( \hat{a}_K^\dagger \right)^n \ket{0}.
\end{align}
The state $\ket{n_1,n_2,...,n_N} = \ket{\mathbf{n}}$ is expressed by the creation operators associated with the pixels, as well as the impinging quantum state $\ket{\psi}$ is described best by means of the creation operator of its mode $K$, so that we get
\begin{align}
\braket{\textbf{n}|\psi} = \frac{1}{(n_1! ... n_N!)^{1/2}} \sum_n \frac{\psi_n}{\sqrt{n!}} \braket{0|\hat{a}_1^{n_1} ... \hat{a}_N^{n_N} (\hat{a}_K^\dagger)^n|0}. \label{Eqnpsi}
\end{align}
From Eq.\,(\ref{Eqnpsi}) it becomes obvious that we need to investigate relations between the creation and annihilation operators and we determine the following useful quantities:
\begin{align}
[\hat{a}_\nu,\hat{a}_K^\dagger] &= \frac{1}{d} \mathscr{U}_{K\nu},\\
[\hat{a}_\nu,(\hat{a}_K^\dagger)^n] &= \frac{n}{d} \mathscr{U}_{K\nu} (\hat{a}_K^\dagger)^{n-1},\\
(\hat{a}_\nu)^m (\hat{a}_K^\dagger)^n\ket{0} &= \frac{n!}{(n-s)!} \frac{\mathscr{U}_{K\nu}^s}{d^s} \hat{a}_\nu^{m-s} (\hat{a}^\dagger_K)^{n-s}\ket{0},\\
&= \begin{cases} \frac{n!}{d^n} \mathscr{U}_{K\nu}^n \hat{a}_\nu^{m-n}\ket{0} = 0 & \text{for } n<m,\\[4pt]
\frac{n!}{(n-m)! d^m} \mathscr{U}_{K\nu}^m (\hat{a}_K^\dagger)^{n-m}\ket{0}& \text{for } n\geq m.
\end{cases}
\end{align}
Using these results, we can determine $\braket{\textbf{n}|\psi}$ as
\begin{align} \label{PIdealn}
\braket{\textbf{n}|\psi} = \psi_\mathcal{N} \sqrt{\frac{\mathcal{N}!}{n_1! ... n_N!}} \prod_{\nu = 1}^N \left( \frac{\mathscr{U}_{K\nu}}{d} \right)^{n_\nu},
\end{align}
where $\mathcal{N} = n_1+...+n_N$ is the total number of measured photons.\newline
According to Eq.\,(\ref{EqP}), we take the modulus square of Eq.\,(\ref{PIdealn}) and obtain
\begin{align}\label{ProbFunc}
P(\textbf{n}|\psi) = |\psi_\mathcal{N}|^2 \frac{(n_1+ ... +n_N)!}{n_1! ... n_N!} p_1^{n_1} ... p_N^{n_N},
\end{align}
with the probabilities $p_\nu \equiv |\mathscr{U}_{K\nu}|^2/d^2$. The amplitude $\psi_\mathcal{N}$ depends on the impinging quantum state, it may for example take the following values:
\begin{align}\label{PsiN}
\psi_\mathcal{N} = \begin{cases} 1, &\mbox{for a number state} \ket{\mathcal{N}}, \\[10pt]
\exp(-\frac{|\alpha|^2}{2}) \frac{\alpha^\mathcal{N}}{(\mathcal{N}!)^{1/2}},&\mbox{for a coherent state} \ket{\alpha}, \\[10pt]
\frac{(\mathcal{N} \text{sech}s)^{1/2}}{(\mathcal{N}/2)!} \left( -\frac{e^{i\vartheta}}{2} \tanh s \right)^{\mathcal{N}/2}, &\mbox{for a squeezed vacuum} \ket{\zeta},\\[10pt]
\frac{n_\text{th}^{\mathcal{N}/2}}{(1+\mathcal{N})^{(1+\mathcal{N})/2}},  &\mbox{for a thermal state},
\end{cases}
\end{align}
with $\zeta = s \exp(i \vartheta)$ and $\mathcal{N}/2 \in $ Integers for the squeezed vacuum state. For the thermal state, the mean photon number $n_\text{th}$ is given by $n_\text{th} = \left[ e^{\beta \hbar \omega} -1 \right]^{-1}$, with $\beta = (k_\text{B} T)^{-1}$, and $k_\text{B}$ being Boltzmann's constant.\newline
The probability distribution in Eq.\,(\ref{ProbFunc}) is a multinomial-like distribution, as anticipated in the introduction. It provides us with comprehensive information about the measurements of the camera. From this, we may derive for example the fluctuations in the position as well as of the width of the beam. We may take advantage of the fact that it is a multinomial-like distribution by exploiting known relations and properties (see Sec.\,\ref{SecAnalyticTheories}). It also simplifies simulations from a technical point of view, as common programs like Mathematica provide built-in functions for multinomial distributions.
\section{Influence of imperfect detectors on measurements} \label{SecImperfectDet}
\begin{figure}[h]
\centering
    \includegraphics[width=0.7\textwidth]{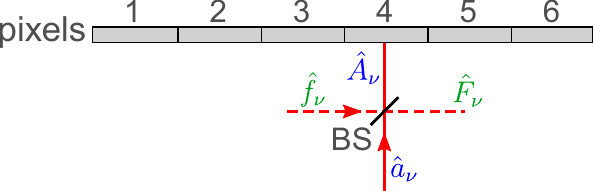}
		\caption{\textit{Modelling an inefficient detector.} To describe an inefficient detector theoretically, one can model it as an ideal detector preceded by beamsplitters, one in front of each pixel. The input signal $\hat{a}_\nu$ under investigation is superimposed with the vacuum $\hat{f}_\nu$ on the beamsplitter (BS). Thus, $\hat{f}_\nu$ is the noise operator that comes into play due to the detection inefficiency. $\hat{A}_\nu$ and $\hat{F}_\nu$ refer to the output modes of the beam splitter. While the mode associated with $\hat{A}_\nu$ is measured by the detector pixel, the mode of $\hat{F}_\nu$ is not measurable in practice.}
    \label{fig:IneffCCD}
\end{figure}

In reality, a photodetector always exhibits imperfections. The dominating effects are a reduced detection efficiency and dark counts. In the following, we discuss how to treat the first. The latter will be taken into account in the simulations in Sec.\,(\ref{SimsResults}).\newline
A reduction of the detection efficiency effectively means that losses occur. In quantum optics, losses are typically modeled via a beamsplitter with appropriate transmittivity and reflectivity \cite{Loudon}. In this way, the vacuum fluctuations come into play by entering the system by the unused port of the beamsplitter. For the multipixel detector, each pixel experiences these losses and we thus may model the detection inefficiency by a beamsplitter preceding each pixel as shown in Fig.\,(\ref{fig:IneffCCD}). 
The input operators thus are the signal $\hat{a}_\nu$ and the noise operator $\hat{f}_\nu$ that represents vacuum. The output operators $\hat{A}_\nu$ and $\hat{F}_\nu$ can be expressed in terms of the input operators as
\begin{align}\label{EqA}
\hat{A}_\nu = \tau\,\hat{a}_\nu + \rho\,\hat{f}_\nu,\\
\hat{F}_\nu = \tau\,\hat{f}_\nu + \rho\,\hat{a}_\nu,\label{EqF}
\end{align}
where $\tau$ is the transmittance and $\rho$ the reflectance of the beam splitter. For each pixel of the detector, there is another pixel mode at the other output port of the beam splitter that we cannot measure. Here, this corresponds to the mode $\hat{F}_\nu$. Following the same approach as in the previous section about the ideal photodetector we want to determine the probability distribution of the photons, i.e. the probability that the pixel centered at $\textbf{x}_{i_1 j_1}$ will count $n_1$ photons and will miss $m_1$ photons, the pixel centered at $\textbf{x}_{i_2 j_2}$ will count $n_2$ photons and will miss $m_2$ photons, ... the pixel centered at $\textbf{x}_{i_N j_N}$ will count $n_N$ photons and will miss $m_N$ photons. We may express it as the overlap between the impinging state $\ket{\psi}\ket{\Omega}$, where $\ket{\Omega}$ denotes the vacuum state, and the state $\ket{\textbf{n},\textbf{m}}$ counting the measured and missed photons $\textbf{n} = n_1, n_2, ... n_N$ and $\textbf{m} = m_1, m_2, ... m_N$, and we write
\begin{align}
P(\textbf{n},\textbf{m}|\psi) &= |\braket{\textbf{n},\textbf{m}|\psi}\ket{\Omega}|^2.
\end{align}
The impinging state consists of the state under investigation 
\begin{align}
\ket{\psi}= \sum_{n=0}^{\infty} \frac{\psi_n}{\sqrt{n!}} \left(\hat{a}_K^\dagger\right)^n \ket{0}
\end{align}
and the vacuum state entering the unused ports of the beam splitters in front of the pixels $\ket{\Omega} = \prod_{\nu = 1}^N \ket{\Omega_\nu}$. We write $\ket{\textbf{n},\textbf{m}}$ in terms of the creation and annihilation operators of the output ports of the beam splitters as 
\begin{align}
\ket{\textbf{n},\textbf{m}} = \prod_{\nu = 1}^N \frac{(\hat{A}_\nu^\dagger)^{n_\nu}}{\sqrt{n_\nu!}} \frac{(\hat{F}_\nu^\dagger)^{n_\nu}}{\sqrt{m_\nu!}}\ket{0}\ket{\Omega}. 
\end{align}
We rewrite the output operators $\hat{A}_\nu^\dagger$ and $\hat{F}_\nu^\dagger$ in terms of the input operators $\hat{a}_\nu^\dagger$ and $\hat{f}_\nu^\dagger$ according to Eq.\,(\ref{EqA}) and Eq.\,(\ref{EqF}). As $\hat{f}_\nu$ and $\hat{a}_\nu$ commute, we may use the binomial theorem and get
\begin{align}
\braket{\textbf{n},\textbf{m}|\psi}\ket{\Omega} = \prod_{\nu = 1}^N \frac{1}{{\sqrt{n_\nu! m_\nu!}}} \Big[ \sum_{k_\nu = 0}^{n_\nu} \sum_{l_\nu = 0}^{m_\nu} \binom{n_\nu}{k_\nu} \binom{m_\nu}{l_\nu} \tau^{n_\nu - k_\nu + l_\nu} \rho^{m_\nu - l_\nu + k_\nu} \nonumber \\ \braket{0| \hat{a}_{\nu}^{n_\nu -k_\nu + m_\nu - l_\nu} |\psi} \braket{\Omega_\nu|\hat{f}_\nu^{k_\nu+l_\nu}|\Omega_\nu} \Big],
\end{align}
where $\braket{\Omega_\nu|\hat{f}_\nu^{k_\nu+l_\nu}|\Omega_\nu}$ is nonzero only if $k_\nu = l_\nu =0$ for all $\nu = 1,2,...,N$. This simplifies the expression significantly as only the first term remains. A straightforward calculation along the lines of the one in the previous section gives
\begin{align} \label{ProbFuncIneff}
P(\textbf{n},\textbf{m}|\psi) = |\psi_\mathcal{D}|^2 \mathcal{D}! \prod_{\nu = 1}^N \frac{T_\nu^{n_\nu}}{n_\nu!} \frac{R_\nu^{n_\nu}}{m_\nu!},
\end{align}
with $T_\nu \equiv |\tau|^2 \mathscr{U}_{K \nu}^2/d^2$ and $R_\nu \equiv |\rho|^2 \mathscr{U}_{K \nu}^2/d^2$ and \mbox{$\mathcal{D} = \mathcal{N} +\mathcal{M}$}, where $\mathcal{N} = n_1 + n_2 + ... + n_N$ and \mbox{$\mathcal{M} = m_1 + m_2 + ... + m_N$}.\newline
In practice, we do not have any information about the missed photons $m_\nu$ as they are lost due to the detector inefficiencies. Consequently, we should trace with respect to the unobserved ports of the beamsplitters and in that way calculate the probability $P(\textbf{n}|\psi)$ of the detection of $\mathcal{N}$ photons. In this way, we obtain
\begin{align}\label{Probreal}
P(\textbf{n}|\psi) = \sum_{m_1 = 0}^{\infty} \sum_{m_2 = 0}^{\infty} ... \sum_{m_N = 0}^{\infty} P(\textbf{n},\textbf{m}|\psi).
\end{align}
We have thus obtained a probability distribution for the photons counted by these individual pixels of a CCD camera with a reduced detection efficiency. The present formula is a multi-pixel generalization of the Kelley-Kleiner formula \cite{KK}. We will use it, as well as the one for the ideal detector derived in the previous section, for simulations of the measurement process. In Sec.\,\ref{SimsResults}, we will explain the approach in detail and show the results of our simulations for the noise in the studied spatial beam parameters.

\section{Analytic theory on the noise in width and position of the light beam} \label{SecAnalyticTheories}
Once the probability distribution for the photon counts is known, one may procede in different ways. In Sec.\,\ref{SimsResults}, we present simulations. In the present section, we want to show how to approach the problem analytically using the properties of the multinomial distribution we derived above. We study the uncertainty in the beam position and beam width. 
The uncertainty in the beam width has been investigated analytically in \cite{BWU} already, yet in an entirely different manner. In the present work, we assume a multipixel detector and, in particular, take the discretization due to the pixelized detector into account. The problem is thus approached from a rather experimental point of view. In \cite{BWU}, a more fundamental concept is used that is independent of the detection system. However, we show that the results of the calculations in \cite{BWU} and in our work utilizing a discretized measurement scheme and the properties of a multinomial distribution are consistent.\newline
As claimed before, we want to approach the problem from a practical point of view. To measure the variance of a quantity in an experiment, one carries out the measurement repeatedly and then determines the variance from the results of these repeated measurements. To attain good statistics, the number of measurements has to be sufficiently large. In the following, we use this procedure to determine the variance of the beam width and the beam position analytically. We start by investigating the beam width noise as this quantity will turn out to exhibit a remarkable behavior.\newline
In every run $r$ of the experiment, we define the beam width as
\begin{align}\label{Wr}
W_r = \frac{1}{\overline{C}} \sum_{\nu = 1}^{N} |\textbf{x}_\nu|^2 n_{\nu r}.
\end{align}
Here $\overline{C} = (1/R) \sum_{r=1}^{R} C_r$ is the average number of counts during the entire experiment, and $C_r = \sum_{\nu = 1}^N n_{\nu r}$ the counts in one run of the experiment. For the moment, we are assuming an ideal detector such that the average number of counts is equal to the average number of photons $\overline{n}$.\newline
The choice of the normalization factor in Eq.\,\ref{Wr} might seem surprising at first sight as $1/\overline{C}$ is used instead of $1/C_r$, which would be standard procedure. We are thus normalizing by the average number of counts during the entire experiment, and not by the average number of counts in the run $r$. To understand the motivation for this practice, let us assume for a moment that we used $C_r$ instead. We would thus write 
\begin{align}
W_r^\prime = \frac{1}{C_r} \sum_{\nu = 1}^{N} |\textbf{x}_\nu|^2 n_{\nu r}.
\end{align}
Accordingly, the average beam width would be determined to be
\begin{align}
\overline{W}^\prime &= \frac{1}{R} \sum_{r=1}^R W_r^\prime\nonumber\\
&= \sum_{\nu = 1}^{N} \left[ |\textbf{x}_\nu|^2 \left( \frac{1}{R} \sum_{r=1}^R \frac{n_{\nu r}}{C_r} \right) \right].
\end{align}
From a theoretical point of view, the average quantity between parentheses in the equation above is not easy to evaluate. For this reason, we use the definition from Eq.\,\ref{Wr} for $W_r$ and determine the average beam width from it as
\begin{align} \label{Wmean}
\overline{W} &= \frac{1}{R} \sum_{r = 1}^{N} W_r,
\end{align}
where $R$ is the total number of runs. By inserting $W_r$ from Eq.\,(\ref{Wr}) into Eq.\,(\ref{Wmean}), we get
\begin{align}\label{EqOverlineW}
\overline{W} = \frac{\displaystyle{\sum_{\nu = 1}^N |\textbf{x}_\nu|^2 \overline{n}_\nu}}{\displaystyle{\sum_{\nu = 1}^N \overline{n}_\nu}},
\end{align}
where $\overline{n}_\nu$ is the average number of photons counted by the pixel $\nu$. To determine the variance $\text{Var}[W] = \overline{W^2}-\overline{W}^2$ of the beam width, we still need to calculate the average value of the squared beam width $\overline{W^2} = 1/R \sum_{r=1}^N W_r^2$. By inserting Eq.\,(\ref{Wr}), we get
\begin{align}\label{EqOverlineWsquared}
\overline{W^2} = \frac{1}{\overline{C}^2} \sum_{\nu,\mu = 1}^N |\textbf{x}_\nu|^2 |\textbf{x}_\mu|^2 \overline{n_\nu n_\mu}.
\end{align}
The fundamental properties of the multinomial distribution \cite{Beyer} that dominates the probability distribution of the photons allows us to determine $\overline{n_\nu}$ and $\overline{n_\nu n_\mu}$ to be equal to
\begin{align}
\overline{n_\nu} &= \sum_{\mathcal{N} = 0}^\infty w_\mathcal{N} E[n_\nu] = \overline{n} p_\nu,\\
\overline{n_\nu n_\mu} &= \sum_{\mathcal{N} = 0}^\infty w_\mathcal{N} E[n_\nu n_\mu] = \overline{n} p_\nu \delta_{\nu \mu} + (\overline{n^2}-\overline{n})p_\nu p_\mu,
\end{align}
with $\overline{n} = \sum_{\mathcal{N} = 0}^\infty w_\mathcal{N} \mathcal{N}$. $E[x]$ denotes the expectation value of the random variable $x$. 
We determine further for $\overline{W}$ from Eq.\,(\ref{EqOverlineW}) and $\overline{W^2}$ from Eq.\,(\ref{EqOverlineWsquared})
\begin{align}
\overline{W} &= \sum_{\nu = 1}^N |\textbf{x}_\nu|^2 p_\nu,\\
\overline{W^2} &= \frac{1}{\overline{n}} \sum_{\nu = 1}^{N} |\textbf{x}_\nu|^4 p_\nu + \frac{\overline{n^2}-\overline{n}}{\overline{n}^2} \left(\sum_{\nu = 1}^N |\textbf{x}_\nu|^2 p_\nu \right)^2.
\end{align}
In analogy to the nomenclature in \cite{BWU}, we use the abbreviations $D \equiv \sum_{\nu = 1}^N |\textbf{x}_\nu|^2 p_\nu$ and $F \equiv \sum_{\nu = 1}^{N} |\textbf{x}_\nu|^4 p_\nu$ in the following, and derive the normalized variance of the beam width as
\begin{align} \label{AnaW}
\frac{\text{Var}[W]}{\overline{W}^2} =& \frac{1}{\overline{n}}\left[\left(\frac{\overline{n^2}-\overline{n}^2}{\overline{n}}-1\right) +\frac{F}{D^2} \right]\\
=& \frac{1}{\overline{n}}\left(Q +\frac{F}{D^2} \right),
\end{align}
where we introduced Mandel's Q parameter \cite{MandelQParameter} in the second step. We use the square of $\overline{W}$ for the normalization to match the dimension of $\text{Var}[W]$. The obtained relation is in perfect accordance with the result for single mode states in \cite{BWU}.\newline
Furthermore, we perform the same calculations for the beam position using both concepts. We define the beam position as the centroid of the spatial intensity distribution. The location of this centroid is given by two coordinates $x_c$ and $y_c$. As we are usually studying symmetric cases, we will restrict ourselves to discussing the $x$-coordinate only. The $y$-coordinate may of course be treated accordingly and for an asymmetric mode, it is conceivable that the noise in the $y$-coordinate is different from the noise in the $x$-coordinate. For one run of the experiment, the $x$-coordinate of the beam position is given by
\begin{align} \label{AnaP}
P_r = \frac{1}{\overline{C}} \sum_{\nu = 1}^N x_\nu n_{\nu_r}.
\end{align}
All the following calculations are performed in perfect analogy to the ones for the beam width and will therefore not be discussed in further detail. As a result, we get for the noise in the beam position
\begin{align} \label{Pnoisediscr}
\text{Var}[P] = \left(D_x-G_x^2\right)\frac{1}{\overline{n}} + \frac{\text{Var}[n]}{\overline{n}^2} G_x^2,
\end{align}
with $D_x \equiv \sum_{\nu = 1}^N x_\nu^2 p_\nu$ and $G_x \equiv \sum_{\nu = 1}^N x_\nu p_\nu$. The aforementioned continuous analytic theory following the approach of \cite{BWU} gives
\begin{align}\label{Pnoisecont}
\langle \delta \hat{P}^2 \rangle= \left(D_{00x}-G_{00x}^2\right)\frac{1}{\langle  \hat{n}\rangle} + \frac{\langle \delta \hat{n}^2\rangle}{\langle  \hat{n}\rangle^2}G_{00x}^2,
\end{align}
with $D_{00x} = \iint x^2 |u_0(x,y)|^2 \text{d}x\text{d}y$ and $G_{00x} = \iint x |u_0(x,y)|^2 \text{d}x\text{d}y$, where $u_0(x,y)$ is the classical mode amplitude of the studied light beam. By comparing Eq.\,(\ref{Pnoisediscr}) and Eq.\,(\ref{Pnoisecont}), one can see that both theories give consistent results again.\newline
We choose and may always choose the coordinate system in such a way that the centroid is located at the origin. In this case, $G_{x}$ is equal to zero. We may thus write $\text{Var}[P] = D_x/\overline{n}$ and see that the noise in the position does not depend on the quantum state of the mean field mode. Please note that this statement is valid for single mode states only. It is in accordance with the findings in \cite{Treps2003, QuNoiseMultipixel2005}.
\section{Measurement procedure, simulations \& results} \label{SimsResults}
The most intuitive and straightforward way of measuring the variance of the beam position or width is to determine the respective beam parameter in several runs as described by Eq.\,(\ref{Wr}) and to calculate the statistical variance from the results of these repeated measurements.\newline
In our simulations we follow the same approach and implement it in Mathematica. By using the determined probability distribution (see Eq.\,(\ref{ProbFunc}) and Eq.\,(\ref{ProbFuncIneff})), we generate numerical values simulating measurement data as one would receive it from the measurement with a CCD camera.\newline
For the simulation, we need to rewrite some of the derived formulas in a way that they are suitable for a computational treatment. To describe the photon probability distribution, we use Eq.\,(\ref{ProbFunc}) and rewrite it as
\begin{align}
P(\mathbf{n}|\psi) = w_\mathcal{N} f(\mathbf{n};\mathcal{N},\mathbf{p}),
\end{align}
where $w_\mathcal{N} = |\psi_\mathcal{N}|^2$ (see Eq.\,(\ref{PsiN})) and $f(\mathbf{n};\mathcal{N},\mathbf{p})$ is the multinomial distribution function in the $N$ variables $\mathbf{n} = n_1, ..., n_N$, with parameters $\mathbf{p} = p_1, ..., p_N$. The multinomial distribution is only valid if at least one photon is impinging on the photodetector. For $\mathcal{N}=0$, we have to assume a discrete uniform distribution in the interval $\{0,0\}$. Accordingly, the total probability distribution for the photodetection of the state $\ket{\psi}$ is given by
\begin{align}\label{Ptot}
P_\text{tot}(\mathbf{n}|\psi) = \sum_{\mathcal{N}=0}^\infty w_\mathcal{N} g(\mathbf{n};\mathcal{N},\mathbf{p}),
\end{align}
where 
\begin{align}
g(\mathbf{n};\mathcal{N},\mathbf{p}) = 
\begin{cases} 
\text{unif}\{0,0\}, & \mathcal{N} = 0, \\
f(\mathbf{n};\mathcal{N},\mathbf{p}), & \mathcal{N} \geq 1.
\end{cases}
\end{align}
This mixed distribution can be simulated by means of a built-in Mathematica function. We describe the implementation in \cite{MathematicaImpl}.\newline
We thus obtain data sets of photon counts for the individual pixels of the photodetector. From these data sets, we are able to determine both the beam position and beam width. By performing a high number of measurement runs, we obtain sufficiently good statistics to determine the variances of these beam parameters.\newline
\begin{figure}[ht]
\centering
    \includegraphics[width=0.6\textwidth]{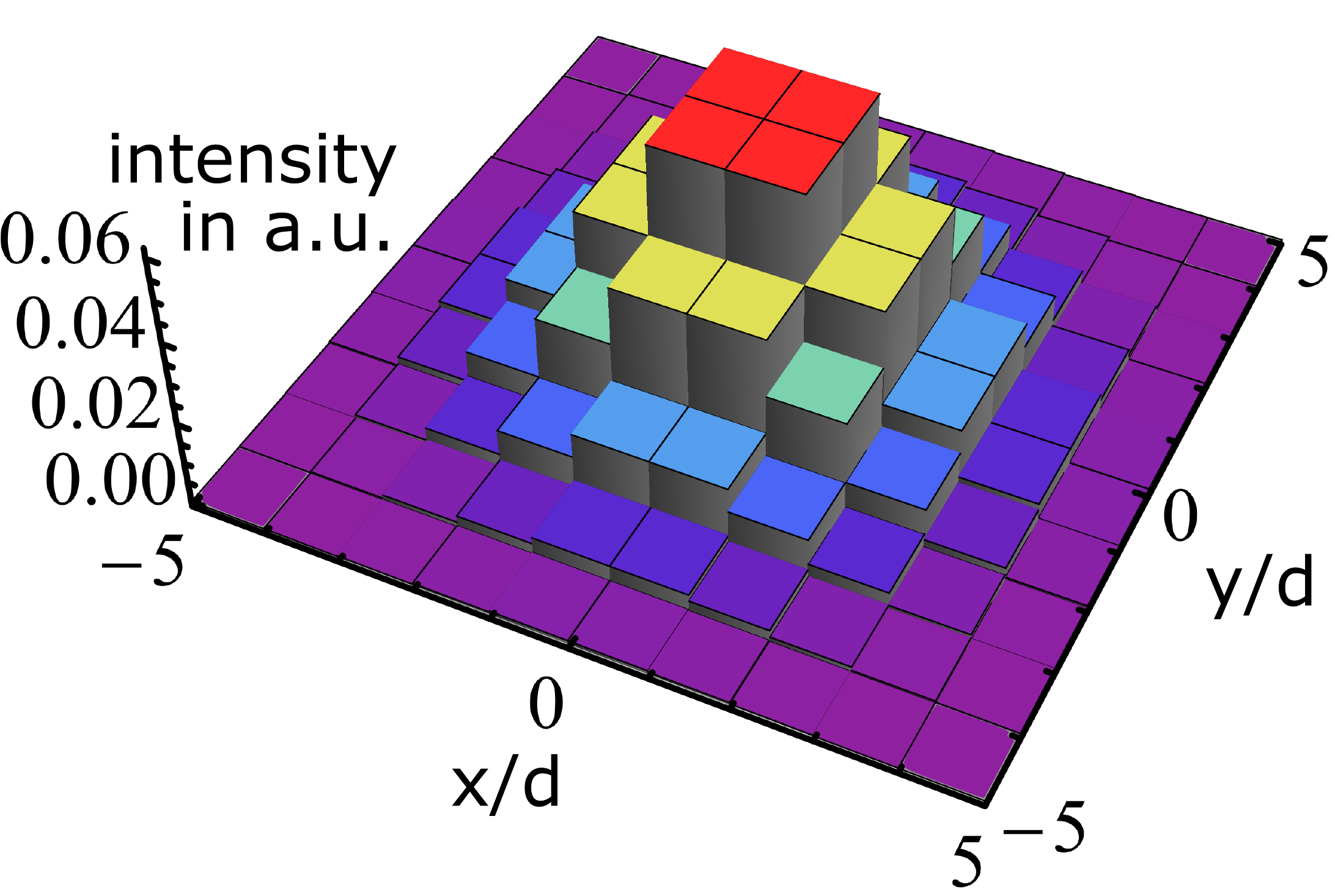}
		\caption{\textit{Fundamental Gaussian beam profile captured by a CCD camera with $10\times10$ pixels.} The discretization of the spatial distribution in the measurement process has to be taken into account: if the number of pixels is too small, an accurate determination of the beam width is not possible.}
    \label{fig:CCDcamSim}
\end{figure}
As an example, we investigate a fundamental Gaussian beam for different quantum states in the following. In particular, we study a coherent, a Fock and a thermal state. For the CCD camera, we assume $10\times10$ pixels. 
Fig.\,(\ref{fig:CCDcamSim}) shows the classical mode shape as it is captured by the CCD camera. We see the discretized shape of a fundamental Gaussian beam profile that covers the CCD camera surface nicely and takes full advantage of the available pixels.
\begin{figure}[h]%
    \subfloat[][]{\includegraphics[width=0.52\textwidth]{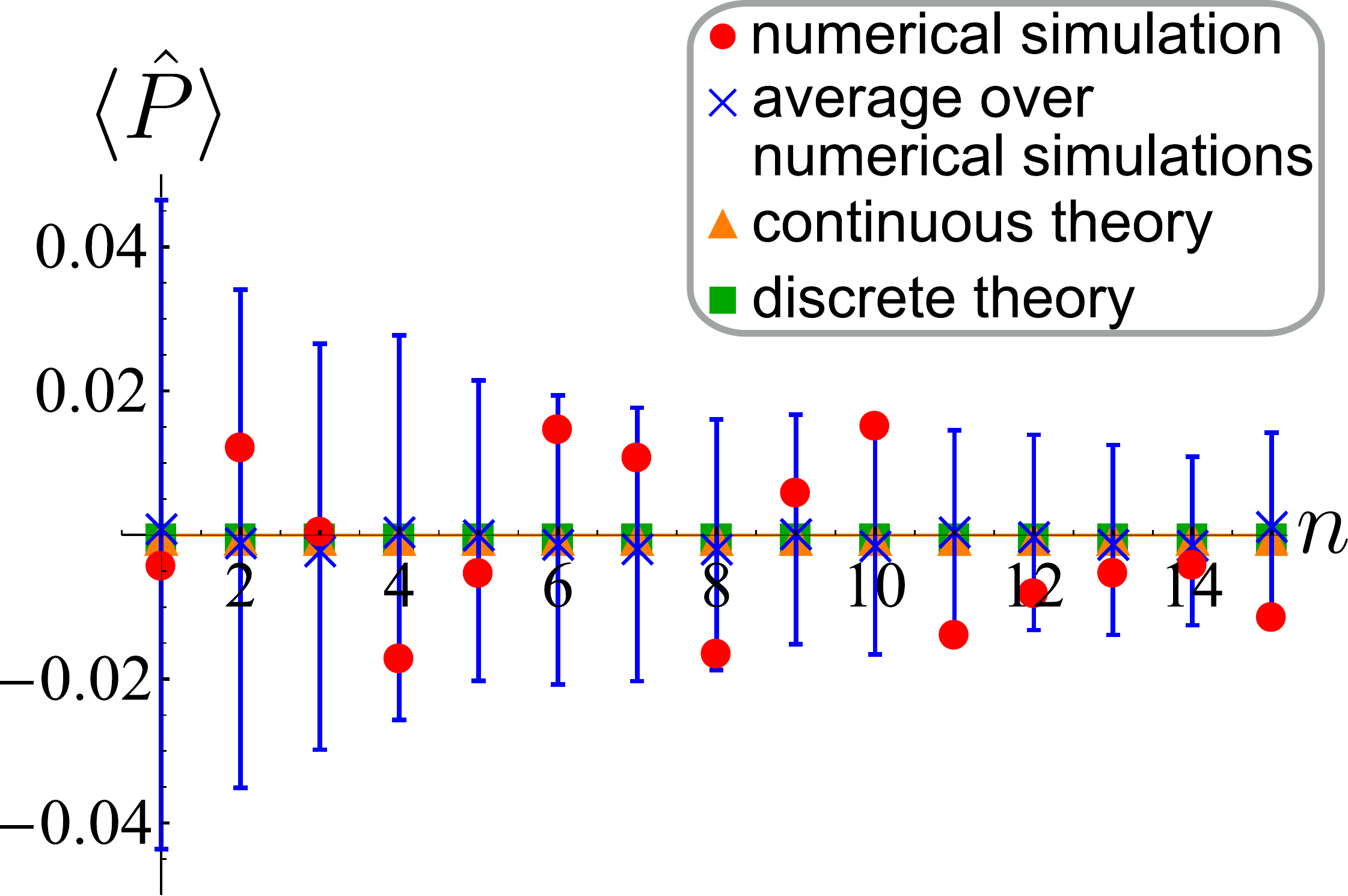}} 
   \hspace{4pt}
   \subfloat[][]{\includegraphics[width=0.48\textwidth]{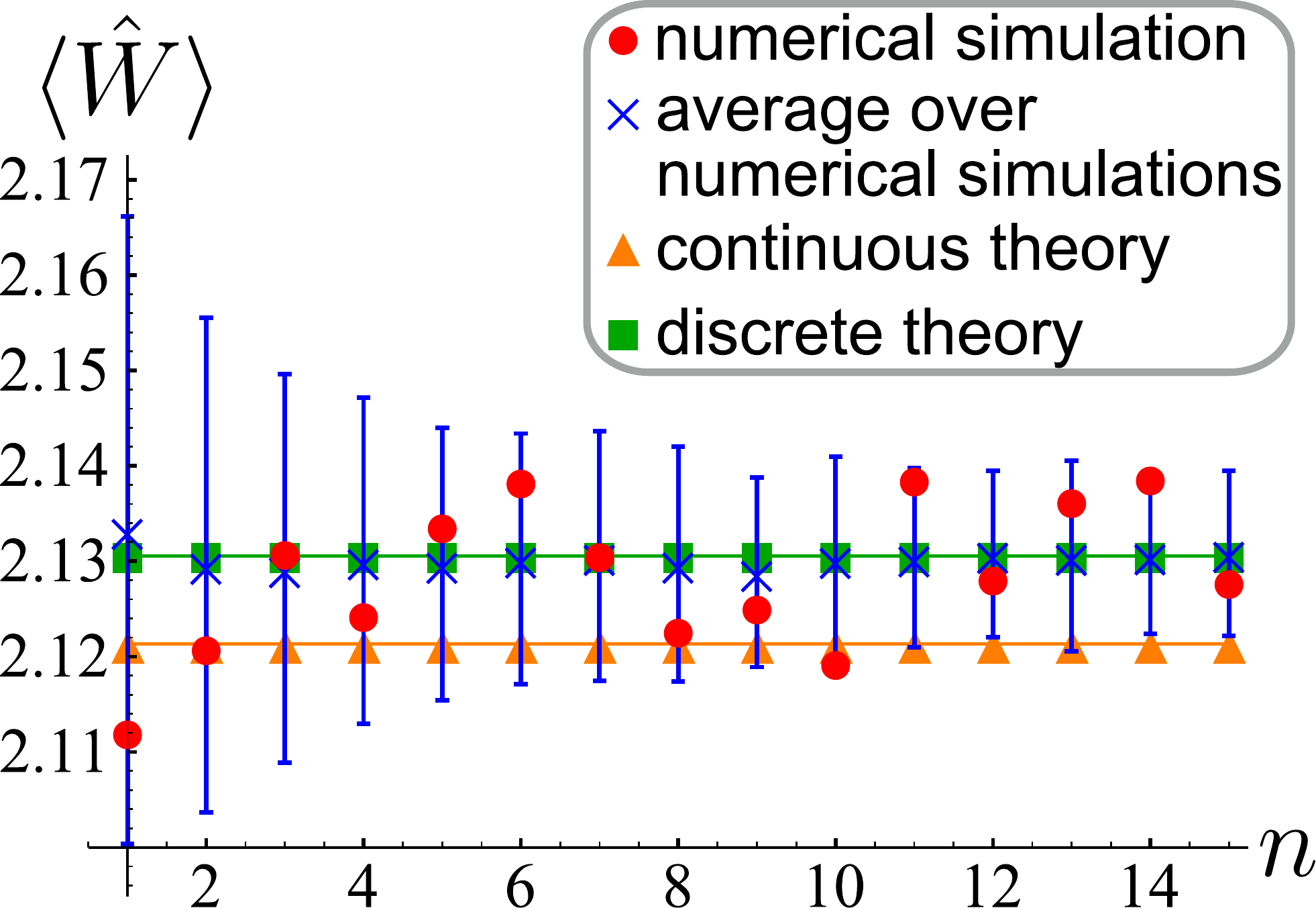}}
    \caption{\textit{Mean values of} (a) \textit{the beam position and} (b) \textit{the beam width for different photon numbers for a coherent state.} The results for the continuous and discrete theory coincide for the beam position, but differ by about 0.5\% for the beam width. This is due to the discretization of the detector that is taken into account for the discrete theory. The results for the numerical simulation (red points) fluctuate by a few percent. These small fluctuations could still be overcome by performing an even higher number of runs $R$ of the experiment. We determined the mean values of the results from 200 numerical simulations and calculated the standard deviation of these such that we obtained the indicated error bars (blue crosses).}
    \label{fig:MeanWP}
\end{figure}
\FloatBarrier
By using the probability distribution in Eq.\,(\ref{Ptot}), we generate data sets. The mean values of beam position and width are depicted in Fig.\,(\ref{fig:MeanWP}). We refer to the theory derived in Sec.\,\ref{SecAnalyticTheories} as discrete theory, while we call the theory presented in \cite{BWU}, which is independent of the measurement scheme, continuous theory in the following. For the beam width noise, we see a difference between the results obtained from the discrete and continuous theory that stems from the discretization of the spatial shape due to the limited number of pixels. The scattering of the data points from the numerical simulation would diminish if a higher number of runs of the experiment was performed. Anyway, these fluctuations amount only to a few percent already. We investigated the reproducibility of the mean values of width and position in the simulations by performing 200 repetitions of the numerical simulation of these values and determined the mean values and the standard deviations that serve as errorbars. 
\begin{figure}[h]%
    \subfloat[][]{\includegraphics[width=0.5\textwidth]{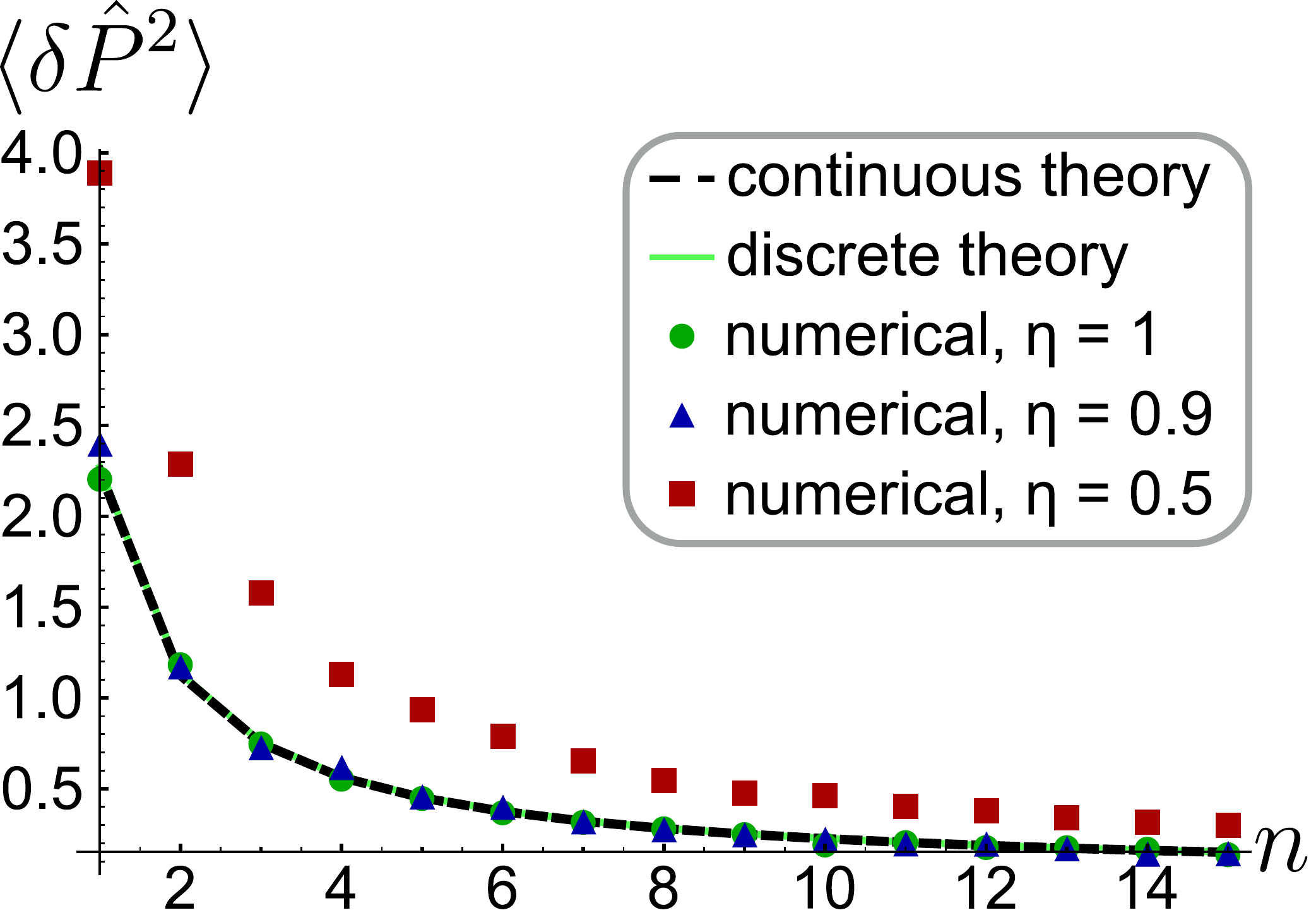}}
   \hfill
   \subfloat[][]{\includegraphics[width=0.5\textwidth]{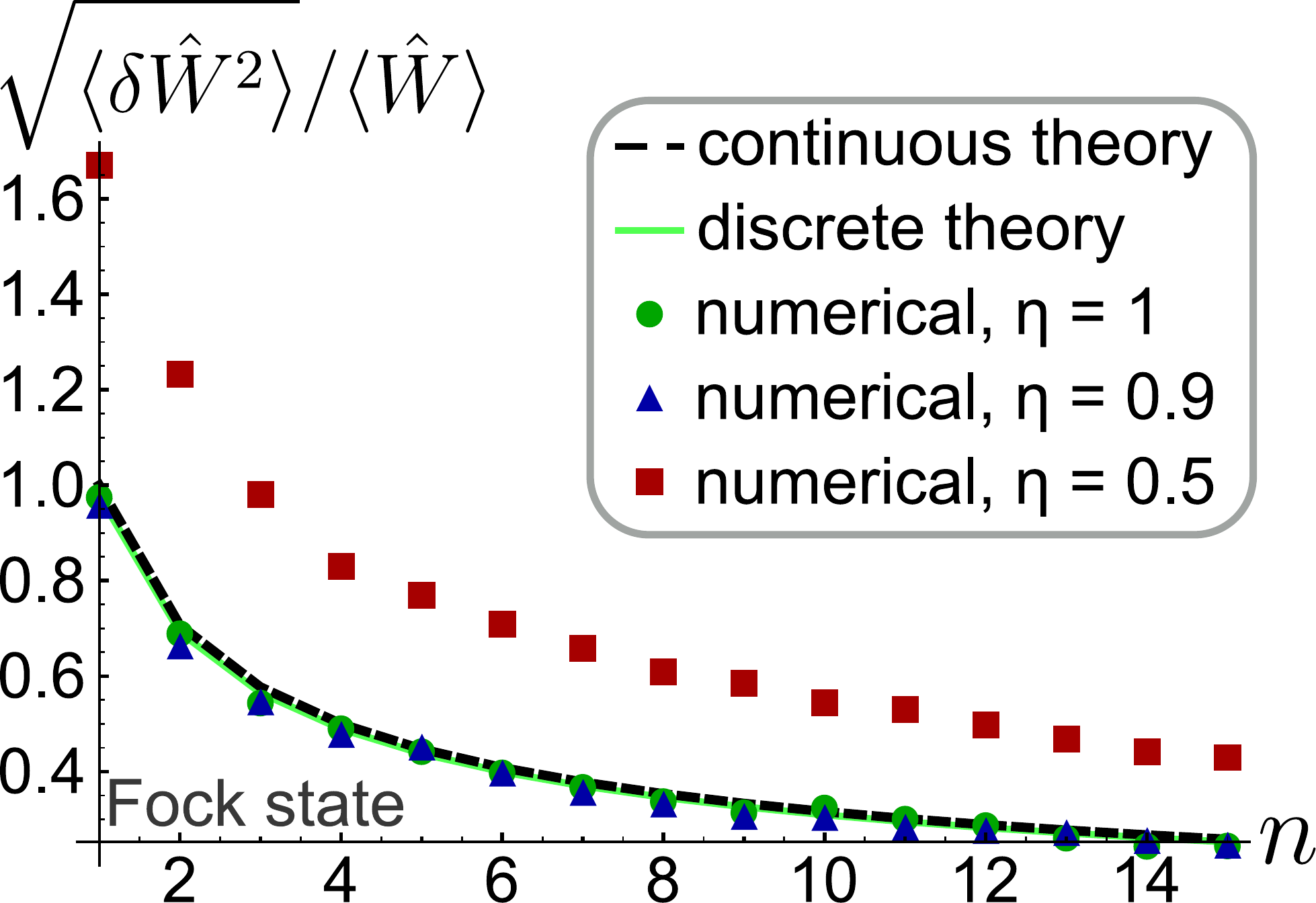}}
    \caption{\textit{(a) Noise in the beam position for different photon numbers (for a Fock state): numerical results, discrete and continuous theory.} Please note that the noise in the beam position is independent of the quantum state, we show exemplarily the plot for a Fock state. From the figure, one can tell that it decreases with increasing photon number. The results from the numerical simulations, discrete and continuous theory for $\eta = 1$ show a good agreement. Even for a reduced detection efficiency of $\eta = 0.9$, the result hardly alters. Whereas, for $\eta = 0.5$ one observes a significant noise increase.\newline
		\textit{(b) Noise in the beam width for different photon numbers for a Fock state: numerical results, discrete and continuous theory.} In general, the noise is decreasing with increasing photon number. The results obtained from the simulations, the discrete and the continuous theory show a good agreement. For a very low detection efficiency of $\eta = 0.5$, the noise in the beam width is on a higher level. This is due to the increased contribution of vacuum fluctuations because of higher losses.}
    \label{fig:Fock}
\end{figure}
\begin{figure}[h]%
    \subfloat[][]{\includegraphics[width=0.5\textwidth]{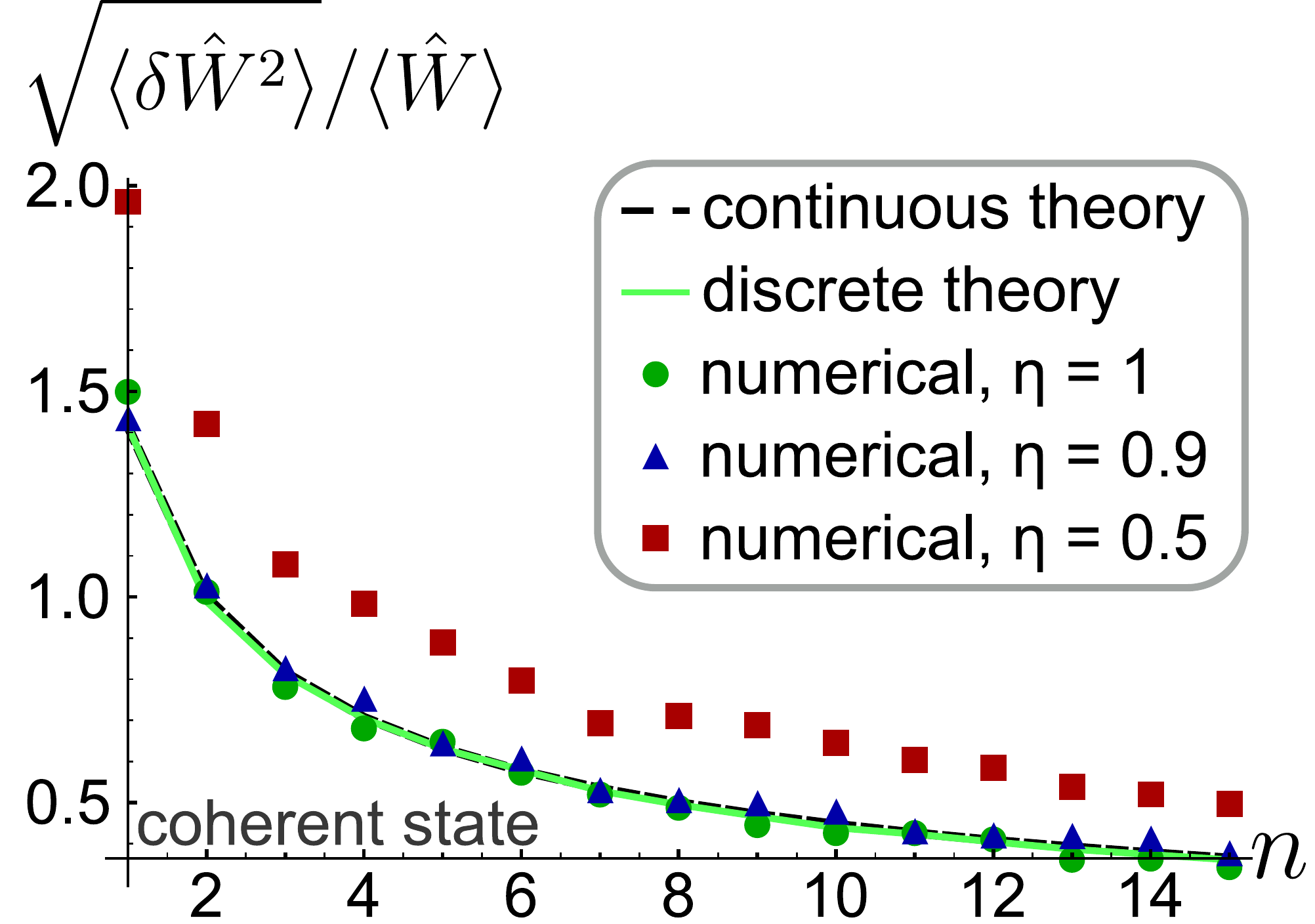}}
   \hfill
   \subfloat[][]{\includegraphics[width=0.5\textwidth]{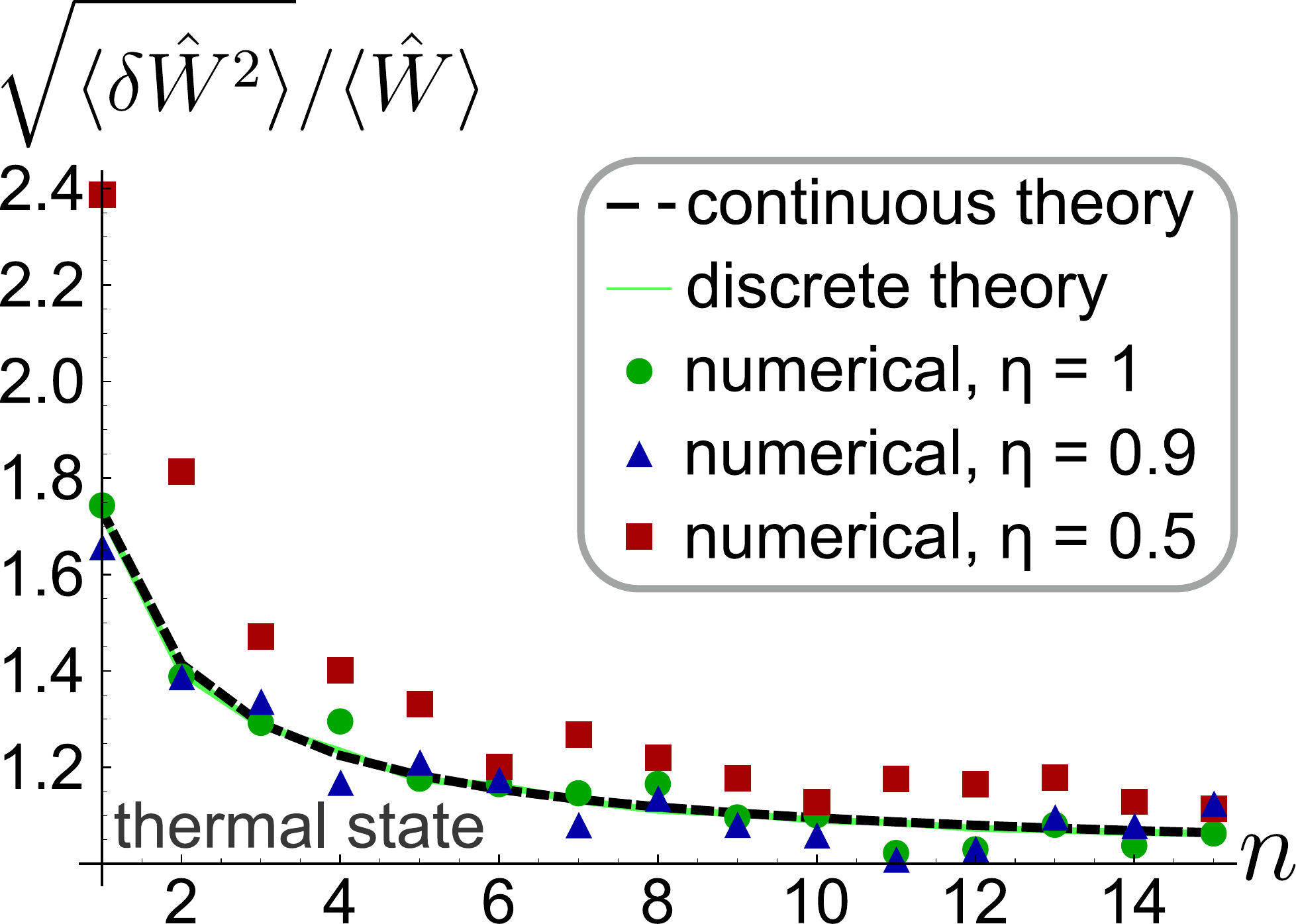}}
    \caption{\textit{(a) Noise in the beam width for different photon numbers for a coherent state: numerical results, discrete and continuous theory.} For the coherent state, the noise level is higher than for the Fock state (see Fig.\,(\ref{fig:Fock}b)). It nevertheless has a similar $\sqrt{1/n}$ dependence. Theory and numerical simulations are in good agreement too.\newline
		\textit{(b) Noise in the beam width for different photon numbers for a thermal state: numerical results, discrete and continuous theory.} For the thermal state, the noise level is higher than for Fock and coherent state. The results of the simulations are less smooth, but still fit to the theory curves.}
    \label{fig:CohThermal}
\end{figure}
Fig.\,(\ref{fig:Fock})-(\ref{fig:CohThermal}) illustrate the noise in beam position and width for coherent, Fock and thermal states as a function of the mean photon number. The numerical results coincide with the results obtained from the discrete theory presented in this article (dashed black line) as well as with the results of the continuous analytic theory (green line) from \cite{BWU}. The green circular data points in the plots indicate the results for a perfect detector. The noise decreases with $1/\sqrt{n}$. For the limit of an infinite photon number, it approaches $0$ for the coherent and the Fock state, and $1$ for the thermal state. The lowest noise level is achieved for the Fock state, the highest for the thermal state. For the blue points, we assumed a reduced detection efficiency of 90\,\% and 10 dark counts per s. The dark counts have been modeled as an additional Poissonian distribution. We choose a measurement time per image of 300\,ns such that the detector exhibits $30\cdot 10^{-9}$ dark counts per image. For the simulations illustrated in Fig.\,(\ref{fig:Fock})-(\ref{fig:CohThermal}), we took $10^3$ images in total. 
It has been proven that these characteristic values can be achieved for a superconducting nanowire single photon detector array in \cite{Nam2013}. One can see that for this detector performance, there is hardly a difference between the results for the perfect and the realistic detector. In order to show what would happen if we used a detector with a very low detection efficiency, we also determined the noise for the beam position and the beam width for a detector with an efficiency of 50\,\%. The results for this case are indicated in red. One can see that for the very low detection efficiency of 50\,\%, the noise increases significantly. Nevertheless, the dependence on the photon number is still perceivable even in this case. \newline
Our results show that a measurement of the noise in beam position and width is feasible by means of state-of-the-art detectors.
\FloatBarrier
\section{Conclusion}
In this article, a method of describing spatial measurements of bosonic particles with multipixel detectors has been presented. We give the probability distribution of the photon counts for the detection of single mode quantum states (Eq.\,(\ref{ProbFunc})), and show explicit expressions for some common states (Eq.\,(\ref{PsiN})). We used this knowledge to investigate the noise in the measurement of spatial beam characteristics, namely beam position and width. Analytic expressions for the beam width noise and the position noise have been derived (see Eq.\,(\ref{AnaW}) and Eq.\,(\ref{AnaP})). We pointed out that the results obtained from the discrete theory derived here coincide with those obtained from the continuous theory put forward in \cite{BWU}.
\newline
Furthermore, we have shown how to take detector inefficiencies into account (Eq.\,(\ref{Probreal})) and included dark counts in our simulations illustrated in Fig.\,(\ref{fig:Fock})-(\ref{fig:CohThermal}). The simulations under realistic measurement conditions give very good results that are extremely close to the theoretical expectations without imperfections.
\newline
We have thus provided a tool for the description of quantum measurements with multipixel detectors and have shown the impact of quantum noise on spatial beam parameters. This can be verified experimentally by means of state-of-the-art detectors.
\FloatBarrier


\begin{thebibliography}{99}
\bibitem{Teich1982} M. C. Teich, P. R. Prucnal, G. Vannucci, M. E. Breton, and W. J. McGill, ``Multiplication Noise in the Human Visual System at Threshold: 1. Quantum Fluctuations and Minimum Detectable Energy,'' J. Opt. Soc. Am. {\bf 72}, 419--431 (1982).

\bibitem{Smith2007} B. J. Smith, and M. G. Raymer, ``Photon wave functions, wave-packet quantization of light, and coherence theory,'' New J. Phys. {\bf 9}, 414 (2007).


\bibitem{BWU} V. Chille, P. Banzer, A. Aiello, G. Leuchs, Ch. Marquardt, N. Treps, and C. Fabre, ``Quantum uncertainty in the beam width of spatial optical modes,'' Opt. Express {\bf 23,} 32777-32787 (2015).

\bibitem{ReviewKolobov} M. I. Kolobov, \lq\lq The spatial behavior of nonclassical light,'' Rev. Mod. Phys. {\bf 71,} 1539--1589 (1999).

\bibitem{QuantumImaging} "Quantum Imaging", M. I. Kolobov, editor (Springer, 2007).

\bibitem{Displacement} C. Fabre, J-B. Fouet, A. Maitre "Quantum limits in the measurement of very small displacements in optical images" Optics Letters, \textbf{25}, 76 (2000).

\bibitem{Resolution} M. Kolobov, C. Fabre, "Quantum limits on optical resolution?, Phys. Rev. Letters  \textbf{85} 3789 (2000).

\bibitem{Gabriel} C. Gabriel, A. Aiello, W. Zhong, T. G. Euser, N. Y. Joly, P. Banzer, M. F\"ortsch, D. Elser, U. L. Andersen, Ch. Marquardt, P. St. J. Russell, and G. Leuchs, \lq\lq Entangling different degrees of freedom by quadrature squeezing cylindrically polarized modes,'' Phys. Rev. Lett. {\bf 106,} 060502 (2011).

\bibitem{Bachor} H.A. Bachor, T.C. Ralph "A guide to experiments in quantum optics" (Wiley, 2004).

\bibitem{Zoller} C. Gardiner, and P. Zoller, \textit{Quantum Noise} (Springer, 2004).

\bibitem{SpatialNoiseLasers1998} J.-Ph. Poizat, T. Chang, O. Ripoll, and Ph. Grangier, ``Spatial quantum noise of laser diodes ,'' J. Opt. Soc. Am. B {\bf 15}(6), 1757 (1998); J.-P. Hermier, A. Bramati, A. Z. Khoury, E. Giacobino, J.-Ph. Poizat, T.J. Chang, and Ph. Grangier, ``Spatial quantum noise of semiconductor lasers,'' J. Opt. Soc. Am. B {\bf 16}(11), 2140 (1999).

\bibitem{Lugiato} L. Lugiato, A. Gatti  "Spatial structure of a squeezed vacuum", Phys. Rev. Letters \textbf{70}, 3868 (1993).

\bibitem{Treps2003} N. Treps, N. Grosse, W. P. Bowen, C. Fabre, H.-A. Bachor, and P. K. Lam, ``A Quantum Laser Pointer,'' Science {\bf 301}, 940 (2003).


\bibitem{Hsu2004} M. T. L. Hsu, V. Delaubert, P. K. Lam, and W. P. Bowen, ``Optimal optical measurement of small displacements,'' J. Opt. B: Quantum Semiclass. Opt.  {\bf 6,} 495-501 (2004).

\bibitem{Barnett} S. Barnett, C. Fabre, A. Maître, "Ultimate quantum limits for resolution of beam displacements", EuroPhys. Journal D, \textbf{22}, 501 (2003).

\bibitem{Delaubert2006} V. Delaubert, N. Treps, M. Lassen, C. C. Harb, C. Fabre, P. K. Lam, and H.-A. Bachor, ``TEM$_{10}$ homodyne detection as an optimal small-displacement and tilt-measurement scheme,'' Phys. Rev. A {\bf 74,} 053823 (2006)

\bibitem{Tsang2015} M. Tsang, ``Quantum limits to optical point-source localization,'' Optica {\bf 2,} 646 (2015).

\bibitem{QuNoiseMultipixel2005} N. Treps, V. Delaubert, A. Ma\^{i}tre, J. M. Courty, and C. Fabre, ``Quantum noise in multipixel image processing,'' Phys. Rev. A {\bf 71,} 013820 (2005).

\bibitem{InPreparation} V. Chille et al., in preparation.

\bibitem{Loudon} R. Loudon, \textit{The Quantum Theory of Light} (Oxford Science Publications, 2000).

\bibitem{MandelWolf} L. Mandel and E. Wolf, \textit{Optical Coherence and Quantum Optics} (Cambridge University Press, 1995)

\bibitem{Bachor2005} H.-A. Bachor, C. Fabre, P. K. Lam, and N. Treps, ``Teaching a laser beam to got straight,'' Contemporary Physics {\bf 46}(6), 295-405 (2005).


\bibitem{Deutsch1991} I. H. Deutsch, ``A basis-independent approach to quantum optics,'' Am. J. Phys. {\bf 59}(9), 834-839 (1991).
\bibitem{Blow1990} K. J. Blow, R. Loudon, S. J. D. Phoenix, and T. J. Shepherd, ``Continuum fields in quantum optics,'' Phys. Rev. A {\bf 42,} 4102 (1990).

\bibitem{Shannon1949} C.E. Shannon, ``Communication in the presence of noise,'' Proc. IRE {\bf 37}(1) (1949).
\bibitem{Greiner1993} W. Greiner, \textit{Quantum Mechanics} (Springer, 1993)
\bibitem{Halcomb1984} L. L. Halcomb, and D. J. Diestler , ``One-dimensional quantum scattering in an eigendifferential basis,'' Am. J. Phys. {\bf 52}(5), 443-445 (1984).
\bibitem{Messiah1961} A. Messiah, \textit{Quantum Mechanics} (Wiley, 1961)

\bibitem{KK} P. L. Kelley, W. H. Kleiner, "Theory of Electromagnetic Field Measurement and Photoelectron Counting", Phys. Rev. 306, \textbf{136}, A 316 (1964).

\bibitem{MathematicaImpl} The mixed distribution and can be written in Mathematica as \mbox{$\text{MixedDistribution}[\{w_0, w_1, ..., w_m\},\{ \text{unif}\{0,0\},$\\$f(\mathbf{n};1,\mathbf{p}), ..., f(\mathbf{n};m,\mathbf{p}) \}]$}, where $m$ is the number of photons that is fixed for a Fock state and unbounded for a coherent state. In the second case, one has to choose an appropriate number for $m$ that is sufficiently large. We determined heuristically that $m = 4\langle n \rangle$ is a reasonable choice.
\bibitem{Beyer} W. H. Beyer, ``CRC Standard Mathematical Tables, 28th ed. Boca Raton,'' FL: CRC Press, p. 532 (1987).
\bibitem{MandelQParameter} L. Mandel, \lq\lq Sub-Poissonian photon statistics in resonance fluorescence,'' Opt. Lett. {\bf 4,} 205--207 (1979).
\bibitem{Nam2013} F. Marsili, V. B. Verma, J. A. Stern, S. Harrington, A. E. Lita, T. Gerrits, I. Vayshenker, B. Baek, M. D. Shaw, R. P. Mirin, and S. W. Nam , ``Detecting single infrared photons with 93\,\% system efficiency,'' Nature Photon. {\bf 7}, 210-214 (2013).
\end{thebibliography}
\end{document}